\documentclass[12pt]{article}

\usepackage{amsmath}
\usepackage{graphics}
\begin{document}
\title{Scattering Length for Lennard-Jones Potentials}

\author{F. J. G\'omez and J. Sesma\thanks{e-mail: javier@unizar.es}\\  \   \\
Departamento de F\'{\i}sica Te\'{o}rica, \\ Facultad de Ciencias,
\\ 50009 Zaragoza, Spain. \\  \ } 

\date{ }                   

\maketitle

\begin{abstract}
 We present an exact procedure that allows one to calculate the scattering length for any potential expressed as an algebraic sum of inverse powers of the inter-atomic distance. We apply it to $(12, s)$ Lennard-Jones potentials, with different values of $s$ . The procedure is suitable for a very precise determination of the intensities of the potential corresponding to zeros or poles of the scattering length.
\end{abstract}


\section{Introduction}

In the study of ultracold dilute gases, the effect of atom-atom collisions is, in most cases, described by means of a regularized
zero-range pseudopotential whose coupling constant is related to the scattering length $a_{\rm scatt}$ of the two-atom system \cite{legg}.
Due to this crucial role, several authors have been concerned to calculate the scattering length for interatomic potentials. Numerical methods are unavoidable in the case of general realistic interactions. There are, however, particular forms of potentials whose scattering length is susceptible
of being computed algebraically. Examples are the hard-core potential with a $r^{-n}$ tail \cite{grib,szmy,gao,kala}, and the potentials discussed by Pade \cite{pade1,pade2}.  These, besides of being analytical at any distance, are able to reproduce the general trend of experimental data.

Motivated by the mentioned papers of Pade, we present here an exact method of computing the scattering length applicable to any sum of inverse power potentials. In particular, to illustrate the method, we consider here a (12, s) Lennard-Jones  type potential
\begin{equation}
V_{LJ}(r)=\frac{\hbar^2\lambda}{2mr_{0}^2}\,\left(\left(\frac{r_{0}}{r}\right)^{12}
- \left(\frac{r_{0}}{r}\right)^s\right), \quad \lambda>0, \quad s>3,   \label{uno}
\end{equation}
where $m$ is the reduced mass of the two interacting atoms, $r_{0}$ represents
the classical turning point, and $\lambda$ is a
dimensionless parameter giving the intensity of the interaction. For the exponent $s$ of the long range part of the potential we have considered three different values: $s=6$, as in the familiar van der Waals potential, $s=4$, as in the case of atom-ion collisions \cite{idzi}, and $s=7$, as in the retarded atom-atom potential \cite{casi}. This last case is exactly solvable, as shown by Pade \cite[Eq. (5) with $n=7$]{pade1}. Nevertheless, as a test, we apply our method to solve it. Our procedure, quite
unconventional, has been successfully applied to the solution of the Schr\"odinger equation with a polynomial potential \cite{gom1,gom2}.
(In fact, the transformation $r\,\Longrightarrow\,1/r$ converts the problems at hand into different anharmonic oscillators.)

\section{Solutions of the Schr\"odinger equation}

The reduced radial Schr\"odinger equation with the potential (\ref{uno}), in the case of energy and angular momentum equal to zero, reads
\begin{equation}
-\,\frac{d^2 R(r)}{dr^2}+\frac{\lambda}{r_0^2}\left(\left(\frac{r_{0}}{r}\right)^{12}
- \left(\frac{r_{0}}{r}\right)^s\right)R(r)=0. \label{dos}
\end{equation}
This differential equation presents two singular points: a regular one at infinity, of indices $-1$ and $0$, and an irregular one at the origin.
Two independent solutions can be immediately written as power series of $r^{-1}$,
\begin{equation}
R_j(r)= \left(\frac{r}{r_0}\right)^{\nu_j}\sum_{n=0}^{\infty}a_{n,j}\,\left(\frac{r}{r_0}\right)^{-n},\quad j=1, 2, \label{tres}
\end{equation}
with
\[
\nu_1=1, \qquad \nu_2=0,
\]
and coefficients $a_{n,j}$ given by the recurrence relation
\begin{equation}
a_{0,j}=1,\quad a_{1,1}=0, \quad a_{n,j}=\frac{\lambda\left(a_{n-10,j}-a_{n-s+2,j}\right)}{(n-\nu_j)(n+1-\nu_j)}. \label{cuatro}
\end{equation}
Those series expansions are convergent for any value of $r$, except for $r=0$. The general solution of Eq. (\ref{dos}) is but a linear combination of the two basic solutions $R_j$. Let us suppose that the (unnormalized) physical solution is given by
\begin{equation}
R_{\rm{phys}}(r)=A\,R_1(r)+B\,R_2(r), \label{cinco}
\end{equation}
where $A$ and $B$ are constants, to be determined, such that $R_{\rm{phys}}$ becomes regular at the origin. Then, for large $r$,
\begin{equation}
R_{\rm{phys}}(r)\sim A\, r/r_{0} + B + o(r^{0}), \label{seis}
\end{equation}
and, consequently, the scattering length would be given by
\begin{equation}
a_{\rm{scatt} } = (-B/A) \,r_{0}\,.  \label{siete}
\end{equation}
The determination of $A$ and $B$ requires the knowledge of the behaviour of $R_1$ and $R_2$ at the origin. This is what is known as the {\em connection problem} for the two singular points of the differential equation (\ref{dos}). In what follows, we present our solution of that problem.

Let us simplify the notation by using, instead of $r$, the dimensionless variable
\begin{equation}
z=r/r_{0}.   \label{ocho}
\end{equation}
In terms of the new variable, the Schr\"odinger equation becomes
\begin{equation}
-z^2\,\frac{d^2w(z)}{dz^2}+\left(\lambda \,z^{-10} -\lambda\, z^{-s+2}\right)w(z) = 0\,,  \label{nueve}
\end{equation}
with
\begin{equation}
 w(z)=R(r).  \label{diez}
\end{equation}
The basic solutions given in Eq. (\ref{tres}) are, in terms of $z$,
\begin{equation}
w_j(z)= z^{\nu_j}\sum_{n=0}^{\infty}a_{n,j}\,z^{-n},\quad j=1, 2 \label{duno}
\end{equation}

In order to represent the behaviour of these functions at the origin, it is convenient to consider the two (formal) Thom\'e solutions at that singular point,
\begin{equation}
w_{\ldots}\sim\exp(\beta_{\ldots} z^{-5}/5)\,z^\mu\,\sum_{n=0}^\infty b_{n,\ldots}\,z^n,  \label{ddos}  \\
\end{equation}
the label $\ _{\ldots}$ being to be replaced by $\ _{\rm{reg}}$ or $\ _{\rm{irr}}$, according to the
behaviour, regular or irregular, of these formal solutions at the origin, which is determined by the sign of $\beta_{\ldots}$.
Substitution of these formal expressions in Eq. (\ref{nueve}) gives the condition $\beta _{\ldots}^2=\lambda$, and consequently
\begin{equation}
\beta_{\rm{reg} }=-\sqrt{\lambda}, \qquad  \beta_{\rm{irr}}=\sqrt{\lambda}, \label{dtres}
\end{equation}
the exponent
\begin{equation}
\mu = \left\{ \begin{array}{ll} 3, & \quad\mbox{for  $s=4, 6$}, \\
3+\beta_{\dots}/2, & \quad\mbox{for $s=7$}, \end{array} \right.  \label{dcuatro}
\end{equation}
and the recurrence relation for the coefficients $b_{n,\ldots}$
\begin{equation}
b_{n,\ldots}= \frac{\lambda\,b_{n+s-7,\ldots}+(n-2)(n-3)\,b_{n-5,\ldots}}{2n\beta_{\ldots}} \label{dcinco}
\end{equation}
in the cases of $s=4$ or $6$, or
\begin{equation}
b_{n,\ldots}= \frac{(n-2+\beta_{\dots}/2)(n-3+\beta_{\dots}/2)\,b_{n-5,\ldots}}{2n\beta_{\ldots}}    \label{dseis}
\end{equation}
if $s=7$. (Notice that in this case the recurrence relation reduces to a linear difference equation of first order and, therefore, it is exactly solvable.)
Then, the behaviour of the basic solutions at the origin can be expressed in the form
\begin{equation}
w_j(z) \sim T_{j,\rm reg}\,w_{\rm reg}(z)+T_{j,\rm irr}\,w_{\rm irr}(z),   \quad \mbox{for} \quad z\to 0.   \label{dsiete}
\end{equation}
The coefficients $T_{j,\rm reg}$ and $T_{j,\rm irr}$ are called {\em connection factors}. Their determination solves the connection problem. It allows also one to obtain immediately the scattering length, as we are going to show. Substitution of the last expressions of $w_j$ in
\begin{equation}
w_{\rm phys}(z) = A\,w_1(z)+B\,w_2(z)   \label{docho}
\end{equation}
gives, for $z\to 0$,
\begin{eqnarray}
w_{\rm phys}(z) \sim  \left(A\,T_{1,\rm reg}+B\,T_{2,\rm reg}\right)\,w_{\rm reg}(z)  \nonumber  \\
+\,\left(A\,T_{1,\rm irr}+B\,T_{2,\rm irr}\right)\,w_{\rm irr}(z),      \label{dnueve}
\end{eqnarray}
Since the physical solution has to be regular at the origin, it must be
\begin{equation}
A\,T_{1,{\rm{irr}}} + B\, T_{2,{\rm{irr}}} = 0, \label{veinte}
\end{equation}
giving for the scattering length
\begin{equation}
a_{\rm{scatt} } = \left(T_{1,{\rm{irr}}}/T_{2,{\rm{irr}}}\right) \,r_{0}\,.  \label{vuno}
\end{equation}
For the computation of the connection factors, we have found convenient to express them as quotients of (constant) Wronskians of the concerned solutions. Taking the Wronskian of both sides of Eq. (\ref{dsiete}) with $w_{\rm reg}(z)$ and isolating $T_{j, \rm irr}$ one obtains
\begin{equation}
T_{j, \rm irr} = \mathcal{W}\left[w_j,w_{\rm{reg}}\right]/\mathcal{W}\left[w_{\rm{irr}},w_{\rm{reg}}\right],  \label{vdos}
\end{equation}
that, substituted in Eq. (\ref{vuno}), gives the scattering length as a quotient of Wronskians,
\begin{equation}
a_{\rm{scatt} } = \left(\mathcal{W}\left[w_1,w_{\rm{reg}}\right]/\mathcal{W}\left[w_2,w_{\rm{reg}}\right]\right) \,r_{0}\,,  \label{vtres}
\end{equation}
A direct computation of the values of these Wronskians written in the form
\[
\mathcal{W}\left[w_j,w_{\rm{reg}}\right] = w_j(z)\,\frac{dw_{\rm{reg}}(z)}{dz} - w_{\rm{reg}}(z)\,\frac{dw_j (z)}{dz}
\]
is to be discarded due to the fact that
the expansion in the definition, Eq. (\ref{ddos}), of $w_{\rm{reg}}$ is an asymptotic one and, in general, does not converge. Our indirect
way of evaluating that kind of Wronskians was explained in Refs. \cite{gom1} and \cite{gom2}, where potentials sum of positive powers of the distance were considered.  For convenience of the reader, we apply, in the next Section, the method to the present case.

\section{Computation of the Wronskians}

Let us introduce the auxiliary functions
\begin{eqnarray}
v_j & \equiv & \exp(-\beta_{\rm{reg}}\,z^{-5}/10)\,w_j  \nonumber \\
 & = & \exp(-\beta_{\rm{reg}}\,z^{-5}/10)\, \sum_{n=0}^{\infty}a_{n,j}\,z^{-n+\nu_j} , \label{vcuatro}  \\
v_{\rm{reg}} & \equiv & \exp(-\beta_{\rm{reg}}\,z^{-5}/10)\,w_{\rm{reg}} \nonumber  \\
 & = & \exp(\beta_{\rm{reg}}\,z^{-5}/10)\, \sum_{n=0}^\infty b_{n,\rm{reg}}\,z^{n+\mu} . \label{vcinco}
\end{eqnarray}
Obviously,
\begin{equation}
\mathcal{W}\left[v_j,v_{\rm{reg}}\right] = \exp(-\beta_{\rm{reg}}\,z^{-5}/5)\, \mathcal{W}\left[w_j,w_{\rm{reg}}\right]  \label{vseis}
\end{equation}
The left hand side of this equation can be written formally, by direct computation, as a doubly infinite sum of powers of $z$,
\begin{equation}
\mathcal{W}\left[v_j,v_{\rm{reg}}\right] \sim \sum_{p=-\infty}^\infty \gamma_{p,j}\,z^{p+\nu_j+\mu},   \label{vsiete}
\end{equation}
with the notation
\begin{eqnarray}
\gamma_{p,j} & = & \sum_{m=0}^\infty b_{m,\rm{reg}} \, \Big(-\beta_{\rm{reg}}\,a_{-p+m-6,j} \nonumber  \\
 & & \ +\ (-p+2m-1+\mu-\nu_j)\,a_{-p+m-1,j}\Big)\,. \label{vocho}
\end{eqnarray}
The value of the $\mathcal{W}[w_\nu,w_{\rm{reg}}]$ can be immediately obtained if we are able to write an expansion of
$\exp(-\beta_{\rm{reg}}\,z^{-5}/5)$, in the right hand side of (\ref{vseis}), with the same powers of $z$ as the expansion
(\ref{vsiete}) of the left hand side. With this purpose, we construct, for each value of $j$, five formal expansions
\begin{equation}
\mathcal{E}_{k,j}(z)=\sum_{n=-\infty}^{\infty}
\frac{\left(-\beta_{\rm{reg}}\,z^{-5}/5\right)^{n+\delta_{k,j}}}
{\Gamma(n+1+\delta_{k,j})}, \quad k=0, 1, \ldots , 4\,.
\label{vnueve}
\end{equation}
of $\exp(-\beta_{\rm{reg}}\,z^{-5}/5)$. Such expansions are
but particular forms of the so called Heaviside's exponential
series \cite{hard}
\begin{equation}
\exp(t)\sim\sum_{n=-\infty}^{\infty}\frac{t^{n+\delta}}{\Gamma(n+1+\delta)}.
\label{treinta}
\end{equation}
For integer $\delta$, it reduces to the familiar series expansion of the exponential function. For any other value of
$\delta$, it is an asymptotic expansion whenever $|\arg (t)|<\pi$.
It becomes evident that, for any set of constants $\{C_{k,j}\}$ ($k=0, 1, \ldots , 4$) satisfying the restriction
\begin{equation}
\sum_{k=0}^{4} C_{k,j} = \mathcal{W}[w_j,w_{\rm{reg}}],  \label{tuno}
\end{equation}
one has from (\ref{vseis})
\begin{equation}
\mathcal{W}[v_j,v_{\rm{reg}}]\sim\sum_{k=0}^{4} C_{k,j}\,\mathcal{E}_{k,j}(z).  \label{tdos}
\end{equation}
By choosing for the  $\delta_{k,j}$ in the expansions $\mathcal{E}_{k,j}$ the values
\begin{equation}
\delta_{k,j}=(-\nu_j-\mu+k)/5,  \label{ttres}
\end{equation}
a comparison of the resulting expansion
in (\ref{tdos}) with that in (\ref{vsiete}) can be done, term by term. One obtains in this way
\begin{equation}
C_{k,j}\,\frac{\left(-\beta_{\rm{reg}}/5\right)^{n+\delta_{k,j}}}
{\Gamma(n+1+\delta_{k,j})} = \gamma_{-5n-k,j}\,, \label{tcuatro}
\end{equation}
for any positive integer $n$.
By substituting in (\ref{tuno}) the values of
$C_{k,j}$ obtained from (\ref{tcuatro}) one has finally
\begin{equation}
\mathcal{W}\left[w_j,w_{\rm{reg} }\right] = \sum_{k=0}^4\frac{\Gamma(n+1+\delta_{k,j})}{(\sqrt{\lambda}/5)^{n+\delta_{k,j}}}
\,\gamma_{-5n-k,j},  \label{tcinco}
\end{equation}
with $\delta_{k,j}$ and $\gamma_{-5n-k,j}$ given, respectively, by (\ref{ttres}) and (\ref{vocho}).

The convergence of series of the type of that appearing in the right-hand side of Eq. (\ref{vocho})  was discussed in Ref. \cite{gom2}.
It is rather slow, namely, like that of the series $\sum_{m=0}^\infty\,2^{-m/5}$. Consequently,
a sufficiently precise arithmetic should be used. The integer $n$ in the right hand side of Eq. (\ref{tcinco}) can be chosen arbitrarily
with the only restriction $n>\sqrt{\lambda}$. In fact, a test of the procedure is the necessary coincidence of the results obtained
for different choices of $n$.

\section{Results}

We show in Fig. 1 the behavior of the values of $a_{\rm{scatt}}$ obtained with our method, for the particular case of $s=6$,
as the intensity $\lambda$ of the potential goes from 0 to 2500.
Notice that we have represented, on the horizontal axis, the values of $\sqrt{\lambda}$ instead of those of $\lambda$. In his study of solvable potentials of the Lennard-Jones type with exponents $(2s-2,s)$, Pade \cite{pade1} has shown that the difference of the values of the square root of the intensity corresponding to two consecutive zeros or poles of $a_{\rm scatt}$ is a constant, namely $2s-4$. In the case of Figure 1, such difference is not exactly, but nearly a constant. Similar graphics are obtained for other values of $s$ in the potential (\ref{uno}). Nevertheless, as far as the sign of the scattering length determines the attractive or repulsive nature of the zero-range pseudopotential and, consequently, the possibility of formation of Bose-Einstein condensates, the precise location of the zeros and poles of $a_{\rm{scatt}}$ is crucial. Moreover, as it is well known, a new bound state arises whenever the increasing intensity of the potential coincides with one of the poles. Our procedure allows to locate zeros and poles, by requiring cancelation of the Wronskians  $\mathcal{W}\left[w_1,w_{\rm{reg}}\right]$ and $\mathcal{W}\left[w_2,w_{\rm{reg}}\right]$ given by Eq. (\ref{tcinco}), with a precision limited only by the number of digits carried in the computation. We report, in Tables 1 and 2, a list of the first ten zeros and poles of the scattering length for the potential (\ref{uno}) with $s=4, 6$ and $7$. The approximate (exact in the case $s=7$)  constancy of the difference of consecutive zeros or poles is manifest.

\begin{figure}
\includegraphics{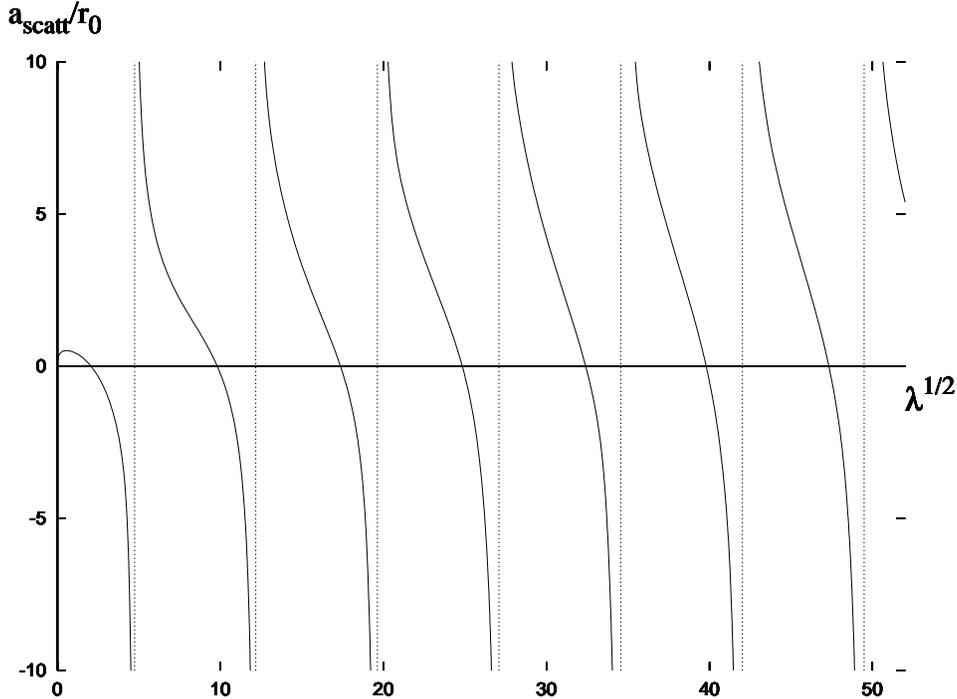}
\caption{Scattering length for the Lennard-Jones potential, Eq. (\ref{uno}) with $s=6$, of intensity $\lambda$ going from 0 to 2500.}
\end{figure}

\begin{table}
 \caption{Values of $\sqrt{\lambda}$ corresponding to zeros of the scattering length of the potential Eq. (\ref{uno}).}
 \hspace{2cm}\begin{tabular}{rrr}
 \hline\noalign{\smallskip}
 $s=4$ \qquad & $s=6$ \qquad & $s=7$ \\
  \hline\noalign{\smallskip}
  1.135708 &  2.944907 &  4 \\
  4.281230 & 10.307414 & 14 \\
  7.627058 & 17.758560 & 24 \\
 10.991652 & 25.220363 & 34 \\
 14.361060 & 32.685259 & 44 \\
 17.732554 & 40.151469 & 54 \\
 21.105133 & 47.618360 & 64 \\
 24.478348 & 55.085650 & 74 \\
 27.851968 & 62.553194 & 84 \\
 31.225862 & 70.020910 & 94 \\
 \noalign{\smallskip}\hline
 \end{tabular}
 \end{table}

\begin{table}
 \caption{Values of $\sqrt{\lambda}$ corresponding to poles of the scattering length of the potential Eq. (\ref{uno}).}
 \hspace{2cm}\begin{tabular}{rrr}
 \hline\noalign{\smallskip}
 $s=4$ \qquad & $s=6$ \qquad & $s=7$ \\
 \hline\noalign{\smallskip}
  2.650141 &  4.728696 &  6 \\
  5.949138 & 12.165518 & 16 \\
  9.308435 & 19.622908 & 26 \\
 12.675992 & 27.086171 & 36 \\
 16.046629 & 34.551611 & 46 \\
 19.418744 & 42.018080 & 56 \\
 22.791679 & 49.485114 & 66 \\
 26.165118 & 56.952491 & 76 \\
 29.538887 & 64.420092 & 86 \\
 32.912885 & 71.887847 & 96 \\
 \noalign{\smallskip}\hline
 \end{tabular}
 \end{table}

 \begin{figure}
\includegraphics{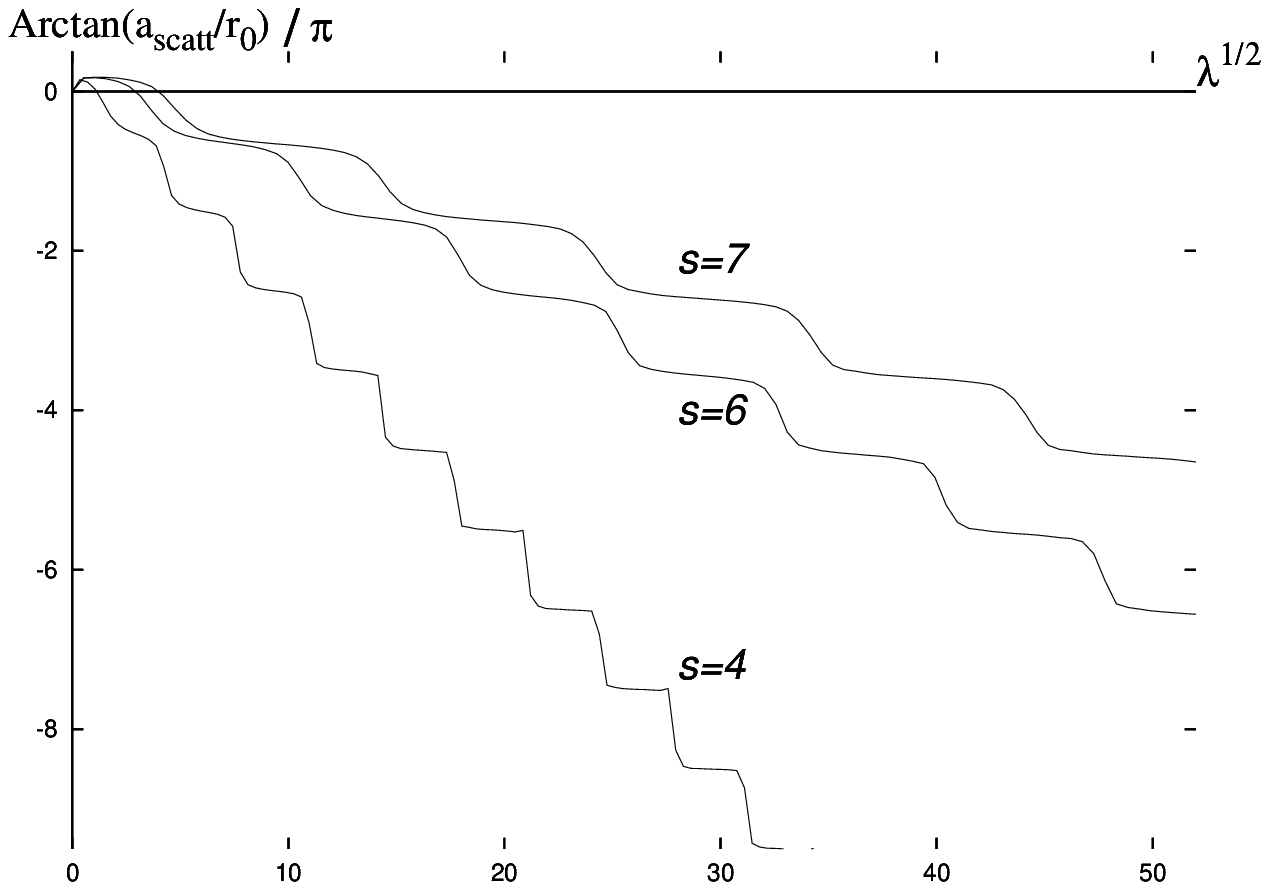}
\caption{An alternative representation of the scattering length of the potential Eq. (\ref{uno}), with different values of the exponent $s$, {\em vs.} the square root of the intensity $\lambda$.}
\end{figure}

 In order to facilitate the comparison of the values of the scattering length for different exponents $s$, we give in Figure 2 a synoptic representation, for $s=4, 6$ and $7$, of a variable related to the scattering length, namely $\arctan(a_{\rm scatt}/r_0)$, {\em vs.} $\sqrt{\lambda}$. Such a variable, that presents the advantage of being free of the singularities of $a_{\rm scatt}$, was already used by Ouerdane {\em et al.} \cite{ouer} to solve numerically the Riccati equation obeyed by the scattering length, in the variable phase theory \cite{calo}, when the potential is truncated at a large distance.

As a consequence of the Pade's study \cite{pade1} and of ours, it seems plausible to formulate the following conjecture: Given a potential
\begin{equation}
V_{LJ}(r)=\frac{\hbar^2\lambda}{2mr_{0}^2}\,\left(\left(\frac{r_{0}}{r}\right)^{s_{\rm rep}}
- \left(\frac{r_{0}}{r}\right)^{s_{\rm attr}}\right), \quad \lambda>0,    \label{tseis}
\end{equation}
of the Lennard-Jones type with exponents $s_{\rm rep} > s_{\rm attr} >3$, let us denote by $\lambda_{0,n}$ and $\lambda_{\infty,n}$ ($n=0,1,2,\ldots$) the values of the intensity corresponding respectively to zeros and poles of the scattering length. Then, one can write the quasi-linear laws
\begin{eqnarray}
\sqrt{\lambda_{0,n}} & = & \mathcal{A}(s_{\rm rep}, s_{\rm attr})\, n + \mathcal{B}_{0,n}(s_{\rm rep}, s_{\rm attr}), \label{tsiete}  \\
\sqrt{\lambda_{\infty,n}} & = & \mathcal{A}(s_{\rm rep}, s_{\rm attr})\, n + \mathcal{B}_{\infty,n}(s_{\rm rep}, s_{\rm attr}), \label{tocho}
\end{eqnarray}
where $\mathcal{B}_{0,n}(s_{\rm rep}, s_{\rm attr})$ and $\mathcal{B}_{\infty,n}(s_{\rm rep}, s_{\rm attr})$ are nearly independent of $n$. In the particular case $s_{\rm rep}=2\,s_{\rm attr}-2$,  Pade \cite{pade1} obtained
\begin{eqnarray}
\mathcal{A}(2\,s_{\rm attr}-2, s_{\rm attr}) & = & 2\, s_{\rm attr}-4, \label{tnueve}  \\
\mathcal{B}_{0,n}(2\,s_{\rm attr}-2, s_{\rm attr}) & = & s_{\rm attr}-3, \label{cuarenta}  \\
\mathcal{B}_{\infty,n}(2\,s_{\rm attr}-2, s_{\rm attr}) & = & s_{\rm attr}-1. \label{cuno}
\end{eqnarray}
The dependence of $\mathcal{A}(s_{\rm rep}, s_{\rm attr})$ on its arguments is such that, keeping $s_{\rm attr}$ fixed, $\mathcal{A}$ diminishes as $s_{\rm rep}$ increases and, with fixed $s_{\rm rep}$, $\mathcal{A}$ increases with $s_{\rm attr}$.

\bigskip

The recommendations of two anonymous referees have been decisive to improve the presentation of this article. The authors acknowledge financial support of Departamento de Ciencia, Tecnolog\'{\i}a y Universidad del Gobierno de Arag\'on and Fondo Social Europeo (Project E24/1) and of Ministerio de Ciencia e Innovaci\'on (Project MTM2009-11154).

\end{document}